\documentclass[
    ,final            
  ]
  {aipproc}

\layoutstyle{6x9}

\newcommand{\id}{{\mathrm d}}
\newcommand{\GeV}{\mbox{GeV}}

\begin{document}

\title{Measurements of $\Delta G$}


\author{Gerhard Mallot}{
  address={CERN, 1211 Geneva, Switzerland}
}


\begin{abstract}
Our present information on the gluon polarisation $\Delta g/g$ is 
reviewed. The data from fixed-target lepton--nucleon experiments are 
in context with the recent data from the RHIC polarised {\sf pp} collider. 
The main tools to study $\Delta g/g$ in lepton--nucleon scattering 
are scaling violations of the $g_1$ structure functions and longitudinal
spin asymmetries in hadron production.
Results from high-$p_{\mathrm T}$ hadron pairs, inclusive hadrons 
as well as open-charm production are discussed. 
At RHIC the most precise data presently came from inclusive $\pi^0$ 
and jet production. 
All data indicate that the gluon polarisation is small compared to
earlier expectations, but still can make a major contribution to the
nucleon spin.
\end{abstract}

\maketitle


\section{Introduction}

Following the discovery \cite{Ashman:1989ig} in 1987 by the European 
Muon Collaboration that the first moment $\Gamma_1$ of spin-dependent 
structure function $g_1$ of the proton is much smaller than expected from the 
Ellis--Jaffe sum rule, the spin structure of the nucleon became
a focus in theoretical and experimental research. By now it is
experimentally well established that indeed the matrix element $a_0$ of the 
flavour-singlet axial-vector current is small and only in the 
order of 0.2--0.3. If the quark spins were responsible for the
nucleon spin a value around 0.6 is expected.
In the $\overline{\mbox{MS}}$ renormalisation scheme the 
sum of the quark spins $\Delta\Sigma=\Delta u+\Delta d+\Delta s$
is given by $a_0$, while in the so-called Adler--Bardeen and JET 
schemes $a_0$ receives an additional contribution from the gluon 
polarisation $\Delta G=\int\Delta g(x,Q^2)\id x$ of 
$-\frac{n_f{\alpha_s}}{2\pi}\Delta G$.
This led to the conjecture
that a large positive $\Delta G$ could explain the smallness of $a_0$
and at the same time reestablish the expected contribution of 0.6 
from the quark spins to the nucleon spin. However, in order to
respect the spin sum rule 
$\frac{1}{2} = \frac{1}{2}\Delta\Sigma+\Delta G + L_{z}$ 
a large gluon polarisation requires partial cancellation 
by orbital angular momentum $L_{z}$. The key to solving the nucleon 
spin puzzle is thus a measurement of gluon polarisation. Now, two 
decades after the original discovery, we hold in hands first data 
suggesting that the large $\Delta G$ scenario is not realised in 
nature.

\section{Scaling violations}

\begin{figure}
  \includegraphics[width=0.9\textwidth]{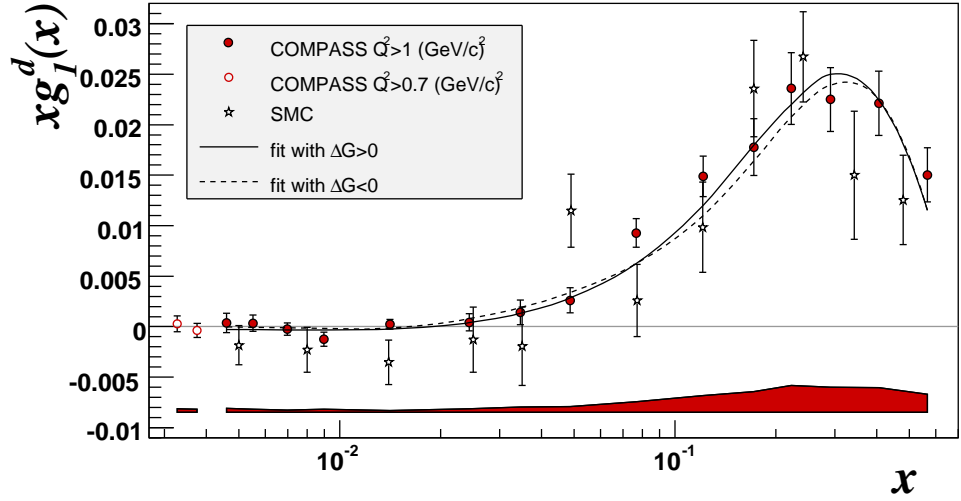}
  \caption{\label{fig:g1d}
     New $xg_1(x,Q^2)$ deuteron data from {\sc Compass} as function of $x$ together 
     with the NLO QCD fit results \cite{Savin_Spin2006}.}
\end{figure}

\begin{figure}
  \includegraphics[width=0.53\textwidth]{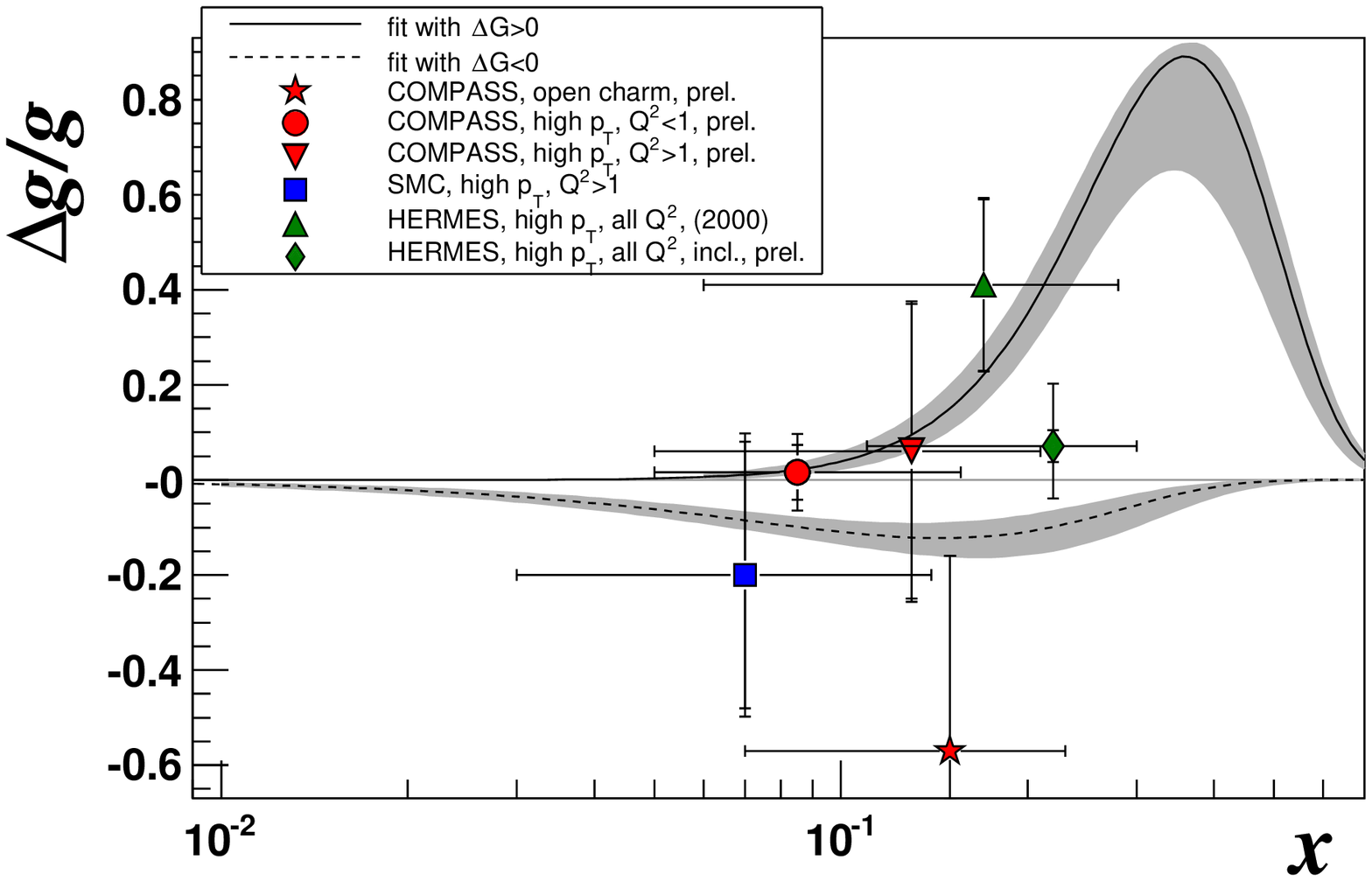}
  \hfill
  \includegraphics[width=0.42\textwidth]{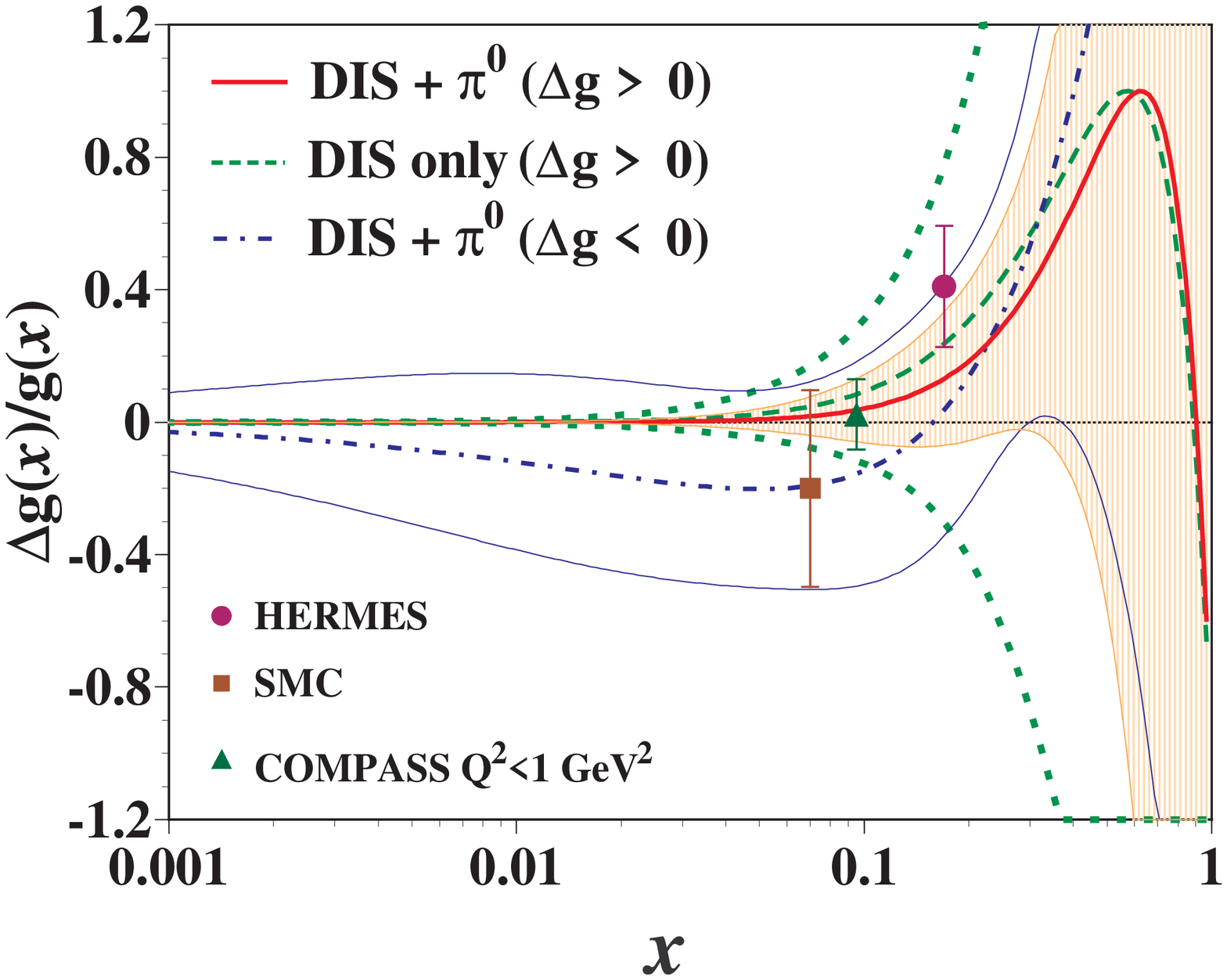}
  \caption{\label{fig:aac_cmp_dg}
     Gluon polarisation $\Delta g/g$ as function of $x$ at $Q^2=Q^2_0$
     obtained by NLO QCD fits (bands) and from LO analysis of hadron 
     helicity asymmetries (symbols).
     {\bf Left:} from {\sc Compass} QCD fits \cite{Savin_Spin2006} including the 
     new {\sc Compass} deuteron data.
     ($Q_0^2=3~\GeV^2$);
     {\bf Right:} from AAC QCD fits \cite{Hirai_Spin2006} involving the 
     $\pi^0$ helicity asymmetries from {\sc Phenix} \cite{Boyle_Spin2006}
     ($Q_0^2=1~\GeV^2$).}
\end{figure}

The $Q^2$ evolution of structure functions is governed by the DGLAP 
equations. Thus the polarised gluon distribution function $\Delta g(x,Q^2)$
can in principle be obtained from a QCD fit to the world data on the
spin-dependent structure function $g_1(x,Q^2)$.
In the unpolarised case, HERA and fixed-target data combined 
offer a huge range in $x$ and $Q^2$. 
Due to the lack of a polarised {\sl ep} collider the situation is 
completely different in the polarised case. 
Here the QCD fits rely largely on the rather small difference in 
$Q^2$ between the SLAC/{\sc Hermes} data on one side and the SMC/{\sc Compass} 
data on the other side, spanning c.m.\ energies from 8~GeV to 20~GeV. 
The status of QCD fits was discussed in detail this 
Symposium \cite{Bluemlein_Spin2006}. 
In the small $x$ region the new deuteron $g_1$ data from 
{\sc Compass} \cite{Savin_Spin2006} are about six times more 
precise than the SMC data. 
{\sc Compass} performed next-to-leading order (NLO) fits to the $g_1$ world 
data including the new deuteron data. 
Two about equally good solutions for $\Delta g(x,Q^2)$ were found, one with a 
positive and one with a negative first moment $\Delta G$ 
(Figs.~\ref{fig:g1d} and \ref{fig:aac_cmp_dg}).
The absolute value is in the order of $|\Delta G|\simeq 0.2 - 0.3$ for 
$Q^2_0=3~\GeV^2$ and the uncertainty from the fit is in the order of 0.1. 
Contributions to the error arising from uncertainties in the factorisation 
and renormalisation scales as well as the influence of the particular 
parametrisation chosen for the distribution functions, were not 
considered. 

Recent fits by the Asymmetry Analysis Collaboration
(AAC) \cite{Hirai_Spin2006} to the world DIS data (including only a 
part of the {\sc Compass} data) also find solutions with negative and 
positive first moments $\Delta G$ at $Q^2_0=1~\GeV^2$. 
In a first step toward a global analysis of data sensitive to $\Delta g$, 
they included the $\pi^0$ helicity asymmetries \cite{Boyle_Spin2006} 
from the {\sc Phenix} 2005 run (see below) in the fit (Fig.~\ref{fig:aac_cmp_dg}). 
A very significant reduction of the uncertainty of $\Delta G$ from
1.08 to 0.32 was observed, at least for the $\Delta G>0$ solution.
However, the sign ambiguity remains. The contribution to $\Delta G$
from the region $x>0.1$ is $0.3$ for both solutions. The negative
first moment for the $\Delta G<0$ solution stems entirely from the
$x<0.1$ region. In contrast to the {\sc Compass} fit, the AAC fit yields 
a gluon distribution with a node around $x=0.16$. 

It should be pointed
out that the two groups apply different error estimations and thus
the error bands shown have different meanings. While the {\sc Compass} 
error band corresponds to the change of $\chi^2$ by unity, 
the AAC bands corresponds to a change of $\chi^2$ by 12.65 for 
their eleven free parameters, what corresponds to 63\% probabilty to
find all parameters simultaneously within one standard deviation rather
than an individual parameter. For details see the documentation
of the {\sc Minuit} program \cite{James:1975dr}.

\section{High-$p_{\sf T}$ hadrons in lepto-production}

To probe the gluon polarisation helicity asymmetries 
were studied for the inclusive production of hadrons, the production
of hadron pairs and of charmed hadrons ({\sl D} mesons). 
For inclusive hadrons \cite{Jager:2005uf} and open-charm \cite{Bojak:1998bd} 
NLO calculations exist, while for hadron pairs NLO is in progress and 
LO is available \cite{Hendlmeier:2006pd}. However, up to now the analyses
were performed in leading order.
Sensitive to the gluon polarisation are the gluon--photon fusion (PGF) 
and the resolved photon parton-level processes. For the latter the 
polarised PDFs of the resolved photon need to be known. The QED part 
can be calculated and the remaining part is bounded by the unpolarised 
PDFs. For hadron-pair production with high transverse momentum $p_{\sf T}$
and $Q^2>1~\GeV^2$ as well as for open-charm production kinematic
regions can be chosen where the PGF process dominates. The longitudinal 
double-spin asymmetry $A_{\parallel}$ is then in leading order 
linear in the gluon polarisation
$$
A_{\parallel} = R_{\sf pgf}
a_{\sf pgf}\frac{\Delta g}{g}+A_{\sf bgd},$$
where $R_{\sf pgf}$ is the fraction of PGF events and 
$a_{\sf pgf}$ is analysing power of the PGF subprocess. 
The latter can be calculated for a given kinematics, but for a particular 
measurement both, $R_{\sf pgf}$ and the average $a_{\sf pgf}$, 
have to be estimated by Monte Carlo (MC) simulations. 
This introduces a model dependence in the 
determination of $\Delta g/g$. 
A possible background
asymmetry $A_{\sf bgd}$ arising e.g.\ from QCD-Compton and direct processes
needs also to be estimated. 
The resulting value of $\Delta g/g$ represents an average of $\Delta g/g(x)$ 
over the probed $x_g$ range, which needs to be determined from the MC simulation. 
The presently available determinations of $\Delta g/g$
from hadron lepto-production are summarised in Table~\ref{tab:dgg} and
shown in Fig.~\ref{fig:aac_cmp_dg}.

\begin{table}
\begin{tabular}{lllllrlr}
\hline
    \tablehead{1}{r}{b}{Experiment}
  & \tablehead{1}{r}{b}{Method}
  & \tablehead{1}{r}{b}{$\Delta g/g$}
  & \tablehead{1}{r}{b}{Stat.\\Error}
  & \tablehead{1}{r}{b}{Sys.\\Error}
  & \tablehead{1}{r}{b}{$\langle \mu^2\rangle$\\$\GeV^2$}
  & \tablehead{1}{r}{b}{$\langle x\rangle$}
  & \tablehead{1}{r}{b}{Published}\\
\hline

{\sc Compass} & hadron pairs ($Q^2<1$) &\hskip8pt0.016 & 0.058 & 0.055                & $\sim3$  & 0.085 & prel., \cite{Ageev:2005pq}\\
{\sc Compass} & hadron pairs ($Q^2>1$) &\hskip8pt0.06  & 0.31  & 0.06                 &          & 0.13  & prel.\\
{\sc Compass} & open charm             &$-0.57$        & 0.41  &                      & 13       & 0.15  & prel.\\
{\sc Hermes}
        & hadron pairs           &\hskip8pt0.41  & 0.18  & 0.03                 & $\sim2$  & 0.17  & \cite{Airapetian:1999ib}\\
{\sc Hermes}  
        & incl. hadrons          &\hskip8pt0.071 & 0.034 & $^{+0.105}_{-0.127}$ & 1.35     & 0.22  & \cite{Liebig_Spin2006}\\
SMC     & hadron pairs ($Q^2>1$) &$-0.20$        & 0.28  & 0.10                 &          & 0.07  & \cite{Adeva:2004dh}\\
\hline
\end{tabular}
\caption{Leading order measurements of $\Delta g/g$}
\label{tab:dgg}
\end{table}

\begin{figure}
\includegraphics[width=0.42\textwidth]{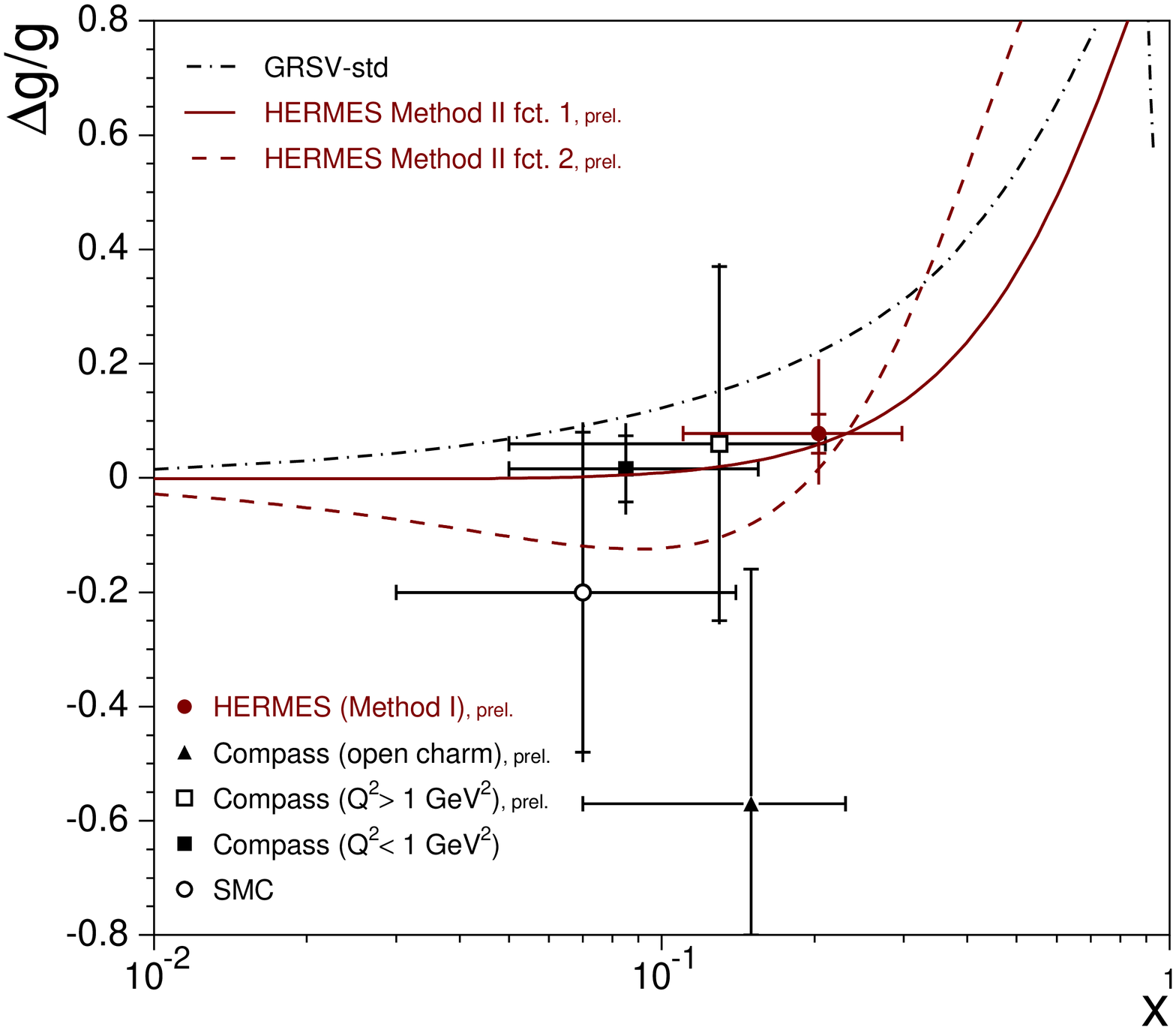}
  \hfill
  \includegraphics[width=0.53\textwidth]{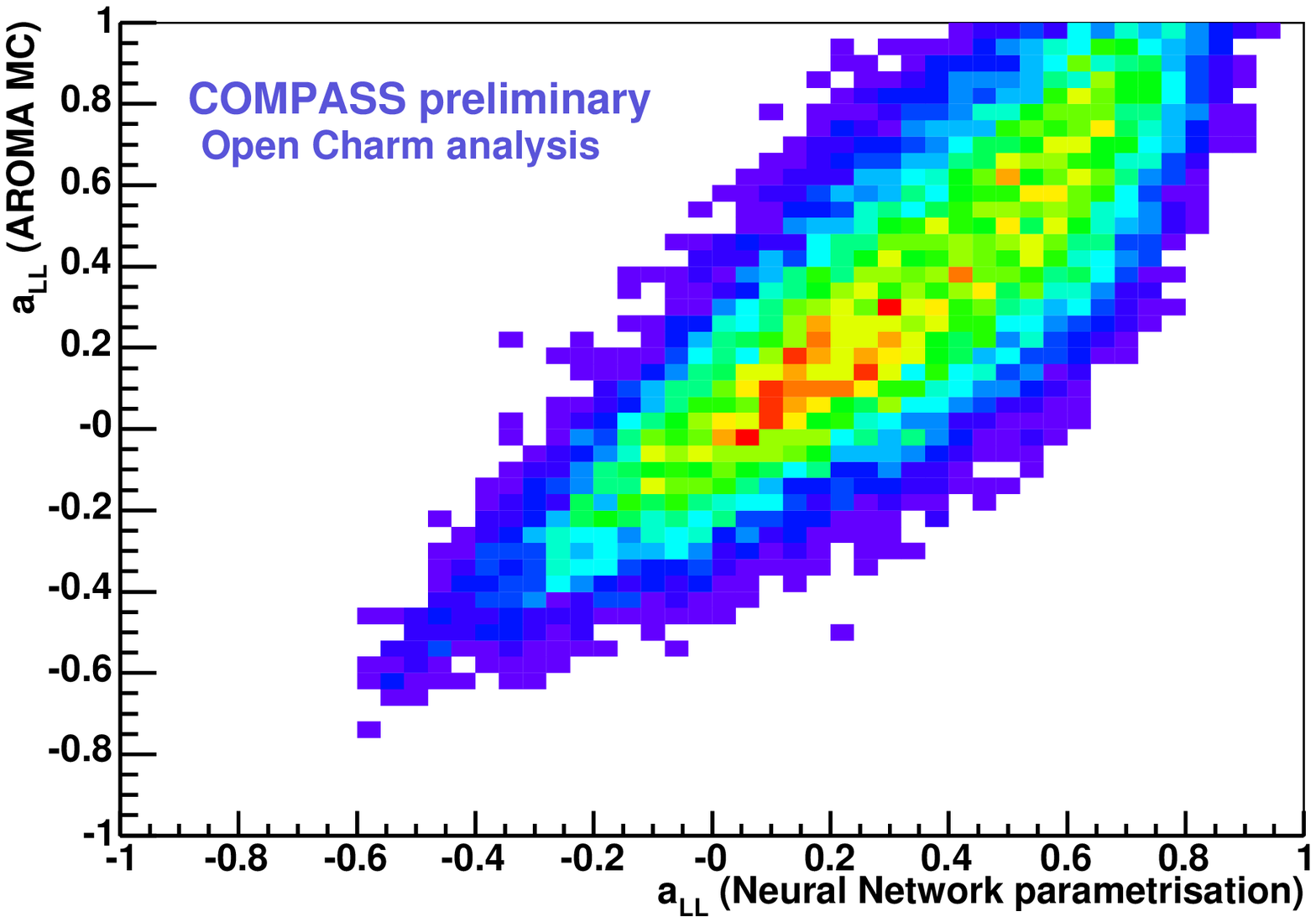}
  \caption{\label{fig:all}
     {\bf Left:} {\sc Hermes} $\Delta g/g$ resulting from fits to data
     in $p_{\sf T}$ bins (Method II). Also shown is a point extracted
     assuming a constant $\Delta g/g$ (Method I).
     {\bf Right:} {\sc Compass} open charm, correlation between the 
     true $a_{\sf PGF}$ from MC and $a_{\sf PGF}$ obtained using a 
     neural network trained on event kinematics.}
\end{figure}

{\sc Compass} performed separate analyses for hadron pairs produced at
$Q^2\ge1~\GeV^2$ and $Q^2<1~\GeV^2$. For both samples transverse momentum
cuts of $p_{\sf T}$ with $p_{\sf T,h1}^2 + p_{\sf T,h2}^2\ge2.5~\GeV^2$ were
applied. The inclusion of the 2004 deuteron data in the low $Q^2$ analysis 
\cite{Ageev:2005pq} yielded an about 1.5 times more precise preliminary result. 
The gluon polarisation from the high and low $Q^2$ analyses are compatible with 
zero and probe the region around $x_g\simeq0.1$. 

{\sc Hermes} presented new analyses of $\Delta g/g$ \cite{Liebig_Spin2006} 
including the deuteron data. Their most precise result comes from inclusive hadron 
asymmetries. A polarised gluon PDF is fitted to $\Delta g/g$ in four $p_{\sf T}$ 
bins in the range $1.05~\GeV<p_{\sf T}<2.5~\GeV$ (`Method II'). 
Figure~\ref{fig:all} (left) shows two of these fits using different functional
forms for $\Delta g/g$. Also shown is a new {\sc Hermes} point for $\Delta g/g$
extracted from the same data, but assuming that $\Delta g/g$ is constant over 
the probed $x_g$ range (Method I). 
The $x_g$ range indicated in Fig.~\ref{fig:aac_cmp_dg} (left) comes from this point
rather than from `Method II'.
The resulting gluon polarisation is again very small and most sensitive to the 
region $x_g\simeq 0.2$. 
The larger 1999 {\sc Hermes} result \cite{Airapetian:1999ib} had been obtained 
from hadron pairs and the data are included in the new analysis. 

Open charm production is considered the most model-independent tool to
study the gluon polarisation. Due to the small initial cross-sections and
the small branching ratio in the decay $D^\circ\rightarrow K\pi$, the 
measurement of an asymmetry in {\sl D} meson production is a real 
challenge.
{\sc Compass} has determined $\Delta g/g$ obtained from this asymmetry. A neural
network was used to estimate $a_{\sf PGF}$ from the kinematics on an
event-by-event basis. The correlation of the network response and the true
$a_{\sf PGF}$ is shown is Fig.~\ref{fig:all} (right). The preliminary result is
smaller than -- but compatible with -- zero (Table~\ref{tab:dgg}, 
Fig.~\ref{fig:aac_cmp_dg} left).

\begin{figure}
  \includegraphics[width=0.45\textwidth]{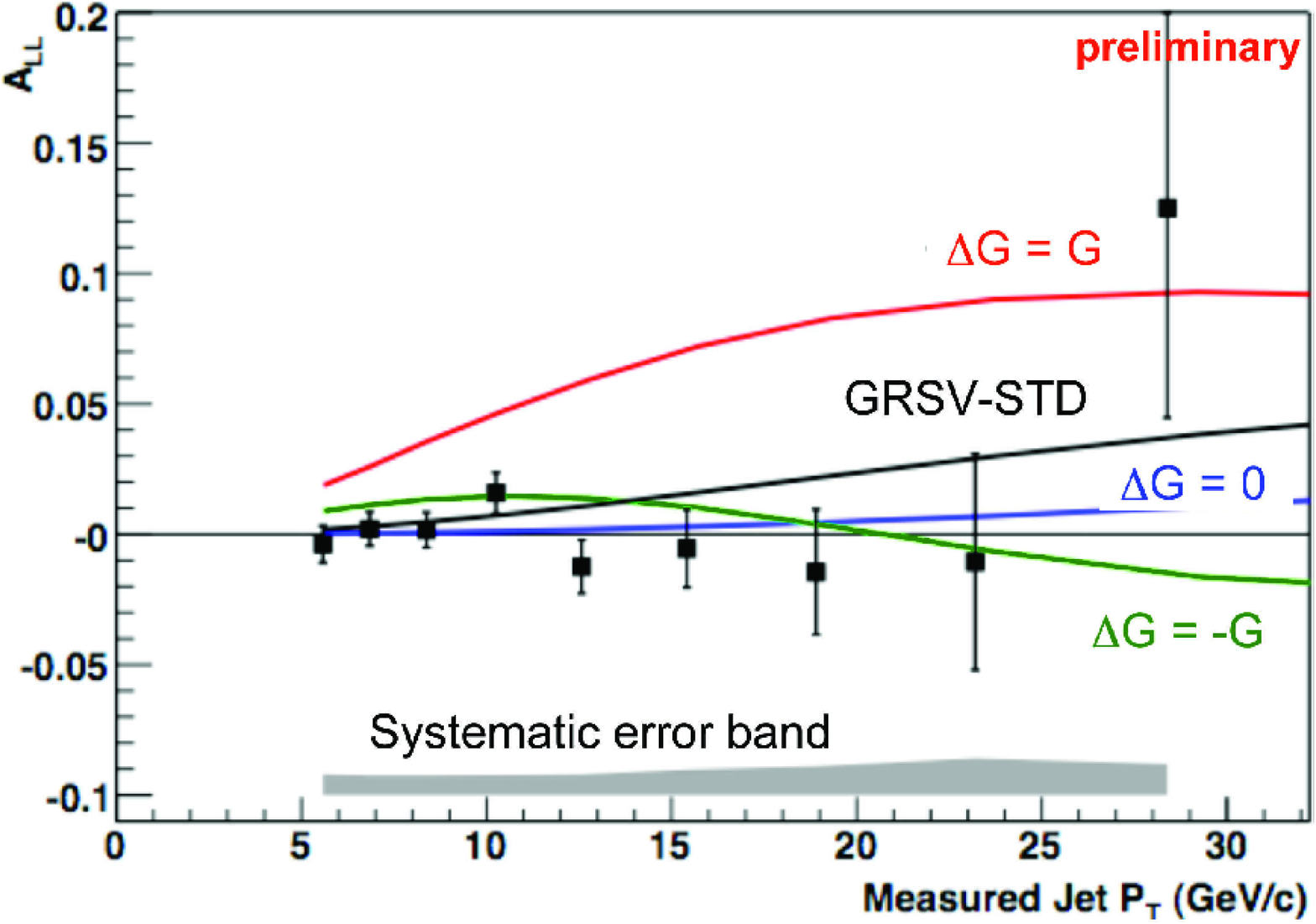}
  \hfill
  \includegraphics[width=0.45\textwidth]{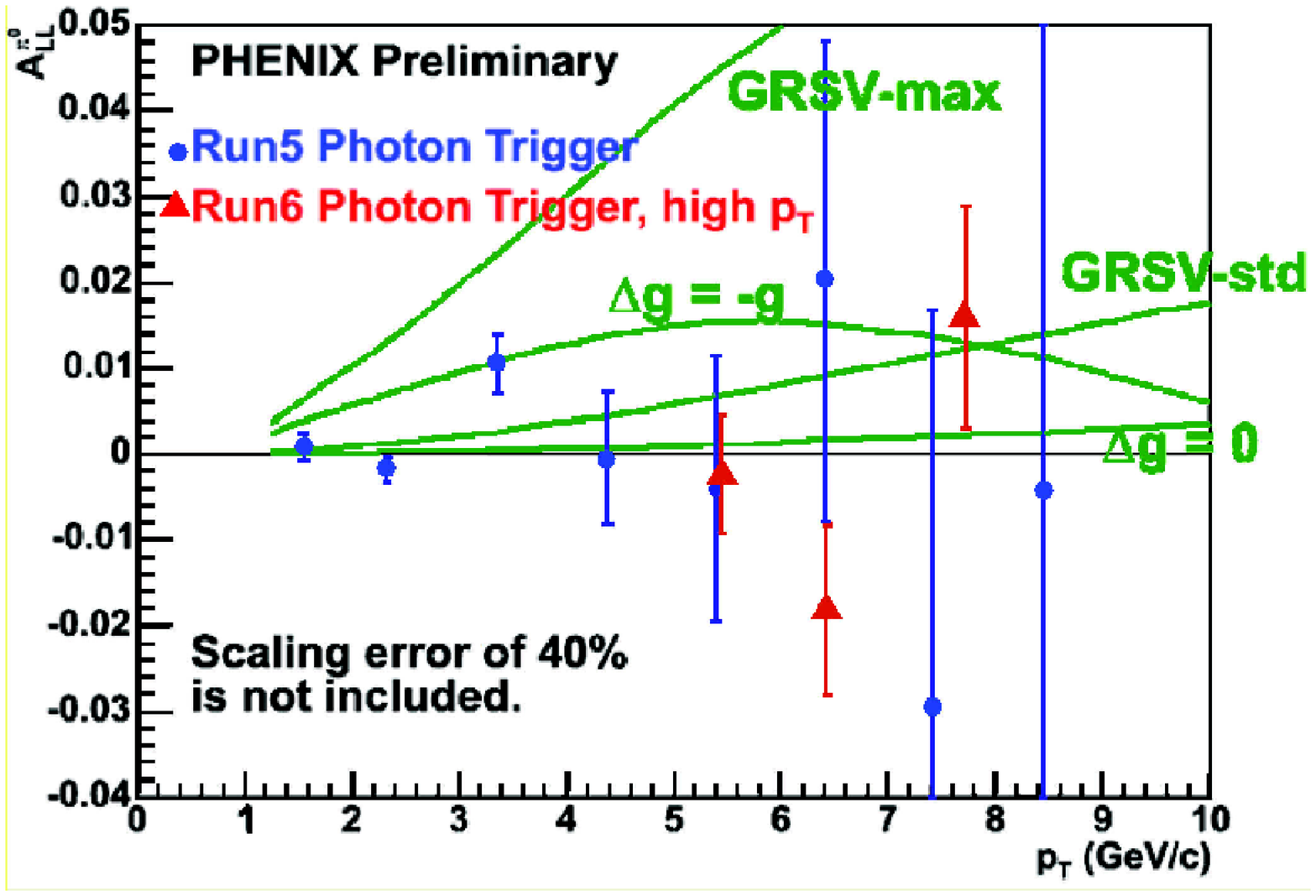}
\caption{\label{fig:rhic} $A_{\sf LL}$ as function of $p_{\sf T}$.
  {\bf Left:}  $A_{\sf LL}$ for inclusive jet production at {\sc Star} \cite{Relyea_Spin2006}.
  {\bf Right:} $A_{\sf LL}$ for inclusive $\pi^0$ production at {\sc Phenix} \cite{Boyle_Spin2006}.
  }
\end{figure}

\section{RHIC data}

The polarised {\sl pp} collider RHIC offers many channels to study the
gluon polarisation. 
At present luminosities and a c.m.\ energy of $\sqrt{s}=200~\GeV$ the most 
promising channels are the inclusive $\pi^0$ \cite{Boyle_Spin2006} and 
jet \cite{Relyea_Spin2006} longitudinal double-spin asymmetries $A_{\sf LL}$
measured by {\sc Phenix} and {\sc Star}. 
The present status of these measurements is shown Fig.~\ref{fig:rhic}. 
Also shown are NLO calculations \cite{Jager:2004jh,Jager:2002xm} using the GRSV 
set of PDFs \cite{Gluck:2000dy} for four different assumptions for the
gluon polarisation, namely the best fit to the world data (GRSV-std) and
$\Delta g = -g$, $0$, $g$ at $Q^2_0=0.3~\GeV^2$. As first observed by
{\sc Compass} \cite{Ageev:2005pq} the data rule out the $\Delta g = g$ scenario,
while the other scenarios are still possible. The dependence of the asymmetry
$A_{\sf LL}$ on $\Delta g/g$ contains a quadratic term which makes it at present 
impossible to determine the sign of $\Delta g/g$ from RHIC data. 
This will improve with the $p_{\sf T}$ range available in the future at 
$\sqrt{s}=500~\GeV$.
Very encouraging are the prospects for the data taken in 2006. 
The average beam polarisation $P_{\sf B}$ improved from 46\% in 2005 to 
62\% and the also luminosity $L$ increased considerably. 
The figure of merit is proportional to $LP^{\,4}_{\sf B}$ and the `Run 6 
high $p_{\sf T}$' {\sc Phenix} data allow us a first glimpse on the 
precision of the new data (Fig.~\ref{fig:rhic} right).
Projections by {\sc Star} \cite{Relyea_Spin2006} for the precision of 
the 2006 jet data are very promising and should allow to distinguish 
between the GRSV-std and the GRSV-min scenarios.

\section{Outlook}

New data from {\sc COMPASS} and RHIC will further clarify the 
composition of the nucleon spin. The restoration of a 60\% 
contribution of the quark spins via the axial anomaly is basically
ruled out.
The new challenge is to precisely determine the fraction of the 
nucleon spin carried by gluon spins rather than distinguishing 
between the various scenarios of the `spin puzzle'.
This must include a precise pinning down of the $x$ dependence
of $\Delta g/g$. For this purpose a global analysis of all relevant
data is indispensable. A particular effort needs to be made to
include the lepton data in a consistent way in the NLO analyses.

The second goal must be a measurement of the angular momentum 
contribution to the nucleon spin. Presently it looks a little bit
like all candidates contribute about equally to the nucleon spin,
a scenario which is particularly hard to prove experimentally.

\begin{theacknowledgments}
I thank the organisers for a wonderful and inspiring Symposium,
P.~Liebig and E.~Aschenauer for interesting discussions on the 
new {\sc Hermes} result, N.~Saito for details on the AAC fit,
and my colleagues from the {\sc Compass} Collaboration for the support 
in selecting and interpreting the material.
\end{theacknowledgments}

\end{document}